\documentclass[aps,pra,twocolumn,showpacs,groupedaddress,superscriptaddress]{revtex4-1}
\usepackage{bm,graphicx}
\usepackage[utf8]{inputenc}
\usepackage[T1]{fontenc}
\usepackage{lmodern}
\usepackage{placeins}
\usepackage{amssymb,amsmath}
\usepackage{epsfig}
\usepackage{natbib}
\usepackage{color}
\usepackage{makecell}
\usepackage{bm}
\usepackage[pdftex,colorlinks,linkcolor=blue,citecolor=blue,urlcolor=blue]{hyperref}
\usepackage{subfigure}
\usepackage{array}
\usepackage{multirow} 
\usepackage{booktabs}
\usepackage{dsfont}
\usepackage[normalem]{ulem}
\usepackage{color}
\usepackage{ulem}
\newcommand{\ket}[1]{\ensuremath{\left| #1 \right\rangle}}

\newcommand{\eg}{{\it{e.g.,~}}}
\newcommand{\ie}{{\it{i.e.,~}}}

\newcommand{\refeq}[1]{Eq.~(\ref{#1})}
\newcommand{\refeqs}[1]{Eqs.~(\ref{#1})}

\newcommand{\1}{\ensuremath{\left|1\right\rangle}}
\newcommand{\2}{\ensuremath{\left|2\right\rangle}}
\newcommand{\3}{\ensuremath{\left|3\right\rangle}}
\newcommand{\4}{\ensuremath{\left|4\right\rangle}}

\newcommand{\D}{\ensuremath{\left|D\right\rangle}}

\newcommand{\comment}[1]{}

\newcommand{\dv}[1]{{\color{blue}#1}}

\definecolor{gray}{gray}{0.6}
\begin{document}
\title{
Atomic\--frequency\--comb quantum memory via piecewise adiabatic passage}
\author{J.~L.~Rubio}\email[]{juanluis.rubio@uab.cat}
\affiliation{Departament de F\'{\i}sica, Universitat Aut\`{o}noma de Barcelona, E-08193 Bellaterra, Spain} 
\author{D.~Viscor}
\affiliation{Departament de F\'{\i}sica, Universitat Aut\`{o}noma de Barcelona, E-08193 Bellaterra, Spain}
\author{J.~Mompart}
\affiliation{Departament de F\'{\i}sica, Universitat Aut\`{o}noma de Barcelona, E-08193 Bellaterra, Spain} 
\author{V.~Ahufinger}
\affiliation{Departament de F\'{\i}sica, Universitat Aut\`{o}noma de Barcelona, E-08193 Bellaterra, Spain}
\date{\today}
\begin{abstract} 
In this work, we propose a method to create an atomic frequency comb (AFC) in hot atomic vapors using the piecewise adiabatic passage (PAP) technique. Due to the Doppler effect, the trains of pulses used for PAP give rise to a velocity-dependent transfer of the atomic population from the initial state to the target one, thus forming a velocity comb whose periodicity depends not only on the repetition rate of the applied pulses but also on the specific atomic transitions considered. We highlight the advantages of using this transfer technique with respect to standard methods and discuss, in particular, its application to store a single telecom photon in an AFC quantum memory using a high density Ba atomic vapor.

\end{abstract}


 
\maketitle

\section{INTRODUCTION}
\label{sec:INTRODUCTION}

The ability to process flying qubits or strings of flying qubits, and specifically the control of light-matter interfaces capable of storing these qubits and retrieve them on demand for subsequent use, \ie quantum memories (QM) \cite{Lvovsky'09}, are key elements for quantum communications \cite{Gisin'07}. Several QM-protocols have been proposed based, for instance, on electromagnetically induced transparency (EIT) \cite{Fleischhauer'00}, controlled reversible inhomogeneous broadening (CRIB) \cite{Moiseev'01,Nilsson'04} and atomic frequency combs (AFC) \cite{Afzelius'09,Afzelius'10,Chaneliere'10,Timoney'12, Riedmatten'08, Zheng'15} (see \cite{Heshami'16} for a review of the most relevant QM protocols), the latter being the most suitable for a multimode storage \cite{Usmani'10,Jobez'16} as the number of modes that can be stored is independent of the optical depth. AFC based QMs have experienced an enormous progress in the last years, achieving spin-wave storage for on-demand retrieval \cite{Gundogan'15,Timoney'12,Yang'18}, high-fidelity multiplexing \cite{Sinclair'14}, optimized efficiencies \cite{Chaneliere'10,Amari'10,Bonarota'10}, and telecom wavelength operation \cite{Gundougan'13,Saglamyurek'15,Jin'15,Farrera'16}.
Typically, the AFC is generated in a static inhomogeneously broadened optical transition in a rare-earth ion doped crystal (REIC) at cryogenic temperatures by means of optical pumping techniques. In this approach, a large number of pulses with temporal spacing $T_{int}$ are repeatedly sent to generate the frequency grating with a detuning spacing $2\pi/T_{int}$. When a signal photon enters the crystal, it is completely absorbed as a single atomic excitation delocalized over the atoms forming the AFC. Subsequently, due to the frequency-to-time conjugation properties, the re-emission of the signal (echo) occurs at a time $T_{int}$.

The ground and the excited states involved to generate the AFC system are usually split in several hyperfine sublevels and one or more of them are used as auxiliary levels for population transfer. Since those different transitions are hidden within the large inhomogeneous broadening, it is first necessary to apply a hole burning (distillation) process in which a wide spectral transmission window is created. Then, the atomic population grating is obtained by optical pumping, which requires many cycles of excitation and de-excitation \cite{Timoney'12}. In other cases, the AFC is directly generated by a combination of $\pi$-pulses that coherently transfer population at different frequencies \cite{Rippe'05}. This approach requires highly accurate control of the pulses intensities. 

Here, we propose an alternative way to produce an AFC using the piecewise adiabatic passage (PAP) \cite{Shapiro'07} technique. PAP is the piecewise version of the well-known stimulated Raman adiabatic passage (STIRAP) \cite{Bergmann'98} technique.
PAP transfers the population between the two internal ground states of an atomic $\Lambda$-system by an accumulative coherent excitation using two trains of pump and dump pulses. We discuss the implementation of this technique in a hot atomic vapor in which a velocity comb (VC) acting as the atomic grating is generated.
Some proposals concerning the generation of VCs by means of a train of short pulses have already been proposed, for instance, to map an optical frequency comb (OFC) into a velocity-selective population transfer between hyperfine levels in Rb \cite{Ban'06}, to achieve Doppler cooling in a two-level atomic system \cite{Ilinova'11}, and to study the coherent control of the accumulative effects in the coherence in a cascade configuration \cite{Felinto'04}.
In our case, the two optical frequency combs of the two PAP trains of pulses selectively transfer the population in a $\Lambda$-type atomic system. Thus, the two-photon resonance condition in a Doppler broadened media is used to generate a velocity comb-like AFC, in order to subsequently store and retrieve a single-photon pulse.
We will consider the mapping of the propagating photon into a storage state different from the ones used for the creation of the velocity comb. The final state where the photon is mapped can be accessed either by a direct one-photon absorption, in which case the retrieval time is predetermined, or by a two-photon process such that the retrieval time can be selected to be the first or any of the subsequent echos. 

An AFC generated via PAP has several advantages: (i) since only a single PAP cycle is needed to complete the transfer of population, the number of required pulses is drastically reduced compared with standard methods \cite{Timoney'12}, (ii) the process exhibits robustness with respect to intensity fluctuations, and (iii) the spacing between the AFC peaks depends not only on the temporal spacing of the PAP pulse train, but also on the ratio between the frequencies of single- and two-photon transitions of the $\Lambda$-type system. Thus, the retrieval time for the QM can be larger compared with other AFC-based QMs generated by a train of pulses with the same temporal spacing.

The article is organized as follows. In Sec \ref{sec:MODEL} we describe the physical system under consideration and the mechanism and conditions to generate an AFC via PAP in an atomic vapor. In Sec. \ref{sec:SIMULATION}, we show numerical simulations for the AFC and use it to investigate the storage and retrieval of a single telecom photon in a Ba atomic vapor. Finally, in Sec. \ref{sec:CONCLUSIONS} we summarize the results and present the conclusions.

\section{PHYSICAL MODEL}
\label{sec:MODEL}

\subsection{Velocity Comb (VC)}
\label{sec:VelocityCombVC}

\begin{figure}[ht]
{
\includegraphics[width=0.9\columnwidth]{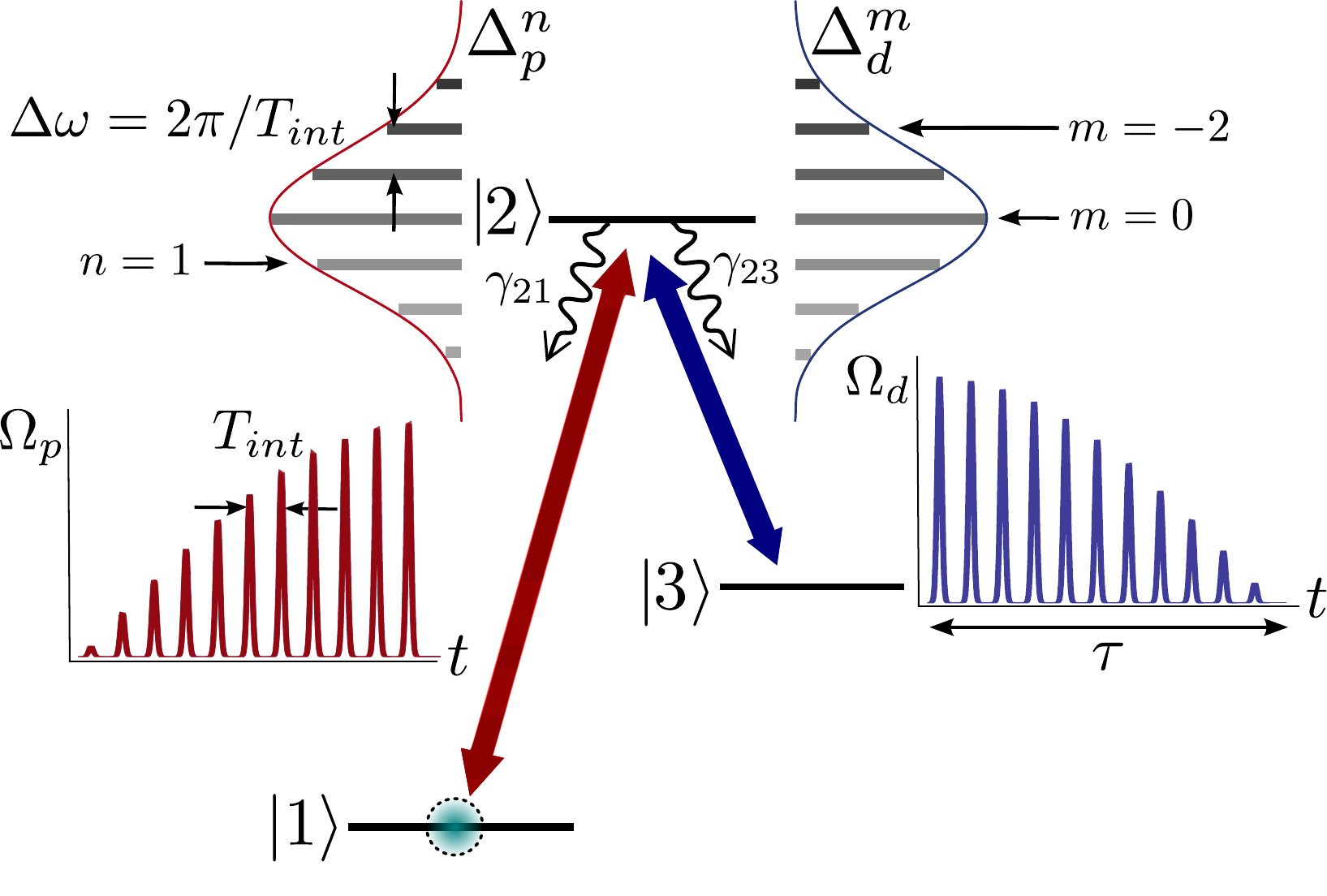}
}
\caption{
Scheme of the $\Lambda$-type system modeling an atom at rest, initially in \1, interacting with the pump and dump trains of pulses with temporal spacing $T_{int}$. In the frequency domain, the fields correspond to an OFC with a frequency separation of $2\pi/T_{int}$ around its nominal frequency. See the main text for the definition of the parameters.
}\label{f:fig1}
\end{figure}

The physical system under consideration consists of an atomic gas in a vapor cell interacting with two co-propagating trains of coherent pulses. The gas is characterized by a Maxwell--Boltzmann velocity distribution,
\begin{equation}
\label{eq:MB}
f(v)=\frac{\varrho}{\sqrt{2\pi}\eta}\exp\left(-\frac{v^2}{2\eta^2}\right),
\end{equation}
where $v$ is the velocity component along the propagation direction of the fields being $v>0\,(v<0)$ for atoms approaching to (moving away from) the fields, $\eta=\sqrt{k_{B}T/m}$ is the velocity standard deviation, $\varrho$ is the total atomic density, 
$m$ is the atomic mass, $k_{B}$ is the Boltzmann constant, and $T$ is the absolute temperature. Each atom of the gas is modeled as a $\Lambda$-type system formed by two ground states \1 and \3, and an excited state \2 (see Fig.~\ref{f:fig1}). The population decay rate from \2 to \1 (\3) is $\gamma_{21}$ ($\gamma_{23}$), and the Bohr frequency of the optical transition \1$\leftrightarrow$\2 (\3$\leftrightarrow$\2) is $\omega_{12}$ ($\omega_{32}$). The atoms are initially in state \1.
A train of $N$ pump (dump) pulses with Rabi frequency $\Omega_{p}(t)=\boldsymbol{\mu}_{12}\cdot\boldsymbol{E}_{p}(t)/\hbar$ [$\Omega_{d}(t)=\boldsymbol{\mu}_{32}\cdot\boldsymbol{E}_{d}(t)/\hbar$] couples the \1$\leftrightarrow$\2 (\3$\leftrightarrow$\2) optical transition with nominal detuning, \ie for atoms at rest, $\Delta_{p}^{0}=\omega_{p}^{0}-\omega_{12}$ ($\Delta_{d}^{0}=\omega_{d}^{0}-\omega_{32}$). Here, $\boldsymbol{\mu}_{12}$ ($\boldsymbol{\mu}_{32}$) is the electric dipole moment vector of the \1$\leftrightarrow$\2 (\3$\leftrightarrow$\2) transition, $\boldsymbol{E}_{p}(t)$ [$\boldsymbol{E}_{d}(t)$] is the slowly varying envelope of the pump (dump) electric field vector while $\omega_{p}^{0}$ ($\omega_{d}^{0}$) is the carrier frequency, and $\hbar$ is the reduced Planck constant.

Due to the Doppler effect, the pump and dump detunings will be shifted by
\begin{subequations}
\label{eq:DopplerDelta}
\begin{align}
\Delta_{p}(v) &= \Delta^{0}_{p} + \omega^{0}_{p}v/c, \\
\Delta_{d}(v) &= \Delta^{0}_{d} + \omega^{0}_{d}v/c,
\end{align}
\end{subequations}
respectively.

Our aim is to create an AFC by means of the PAP technique, which is the piecewise version of STIRAP. In STIRAP, two single pulses, the pump and the dump, couple the two optical transitions of a $\Lambda$-type system, fulfilling the two-photon resonance condition $\Delta^{0}_{p}=\Delta^{0}_{d}$. Under this condition, one of the eigenstates of the Hamiltonian, the so-called dark state, takes the form
\begin{equation}
\label{eq:darkstate}
\D=\cos\theta(t) \1-\sin\theta(t) \3,
\end{equation}
where $\theta(t)=\arctan[\Omega_{p}(t)/\Omega_{d}(t)]$.
Therefore, by smoothly varying the value of $\theta$ from $0$ to $\pi/2$ one can efficiently transfer the atomic population from state \1 to state \3, adiabatically following the dark state. The desired variation of the mixing angle $\theta$ is achieved by coupling the pulses with the atoms in the so called counterintuitive sequence, \ie if the atoms are initially in \1 the sequence consists of applying first the dump and with a certain delay in time $\tau$ the pump. To avoid the coupling between \D\, and the other eigenstates of the system, the required global adiabatic condition reads $\Omega\tau>10\pi/\sqrt{2}$ for optimally delayed Gaussian pulses \cite{Bergmann'98}, where $\Omega^2=\Omega_{p0}^2+\Omega_{d0}^2$, being $\Omega_{p0}$ ($\Omega_{d0}$) the peak value of the pump (dump) Rabi frequency.

In PAP, the pump and the dump pulses are replaced by two trains of pulses separated by an inter-pulse period, $T_{\rm int}$, and with an envelope that follows the temporal sequence of STIRAP (Fig.~\ref{f:fig1}). As it has been previously reported \cite{Shapiro'07,Shapiro'08}, the population is transferred from \1 to \3 by accumulation of the coherence as long as the inter-pulse period is shorter than the atoms decoherence time.
We consider the following temporal profiles for the pulse trains:
\begin{subequations}\label{eq:fields}
\begin{eqnarray}
\Omega_p(t)&=&\Omega_{p0}\,e^{-(t-\tau)^2/2\sigma_{e}^2} \sum_{l=0}^{N-1}\,e^{-(t-l T_{ int})^2/2\sigma^2}, \label{eq:fieldP}\\
\Omega_d(t)&=&\Omega_{d0}\,e^{-t^2/2\sigma_{e}^2} \sum_{l=0}^{N-1}\,e^{-(t-l T_{int})^2/2\sigma^2}. \label{eq:fieldD}
\end{eqnarray}
\end{subequations}
Each of these expressions corresponds to a sum of $N$ narrow Gaussian pulses of width $\sigma$, separated by $T_{int}$, and modulated by a Gaussian envelope of width $\sigma_e$. The pump and dump individual pulses are coincident in time while the corresponding envelopes are shifted with respect to each other by a time $\tau=(N-1)T_{int}$, during which the PAP takes place (see Fig.~\ref{f:fig1}).
In what follows, we will use for simplicity $\Omega_{0}=\Omega_{p0}=\Omega_{d0}$.

Standard STIRAP requires to fulfill the two-photon resonance condition, which for a Doppler-broadened medium is equivalent to set $\Delta_{p}(v)=\Delta_{d}(v)$. However, in the case of a train of pulses one has to look for the accumulation of the coherence between two consecutive pulses. In particular, this effect has been studied in Ref. \cite{Felinto'04} for a cascade configuration. In our scheme, the conditions for constructive interference in the accumulation of the coherences are given by $\Delta_p(v)T_{int}=j2\pi$ and $\Delta_d(v)T_{ int}=k2\pi$, with $j,k \in \mathds{Z}$. Taking the difference between these two expressions, the two-photon resonance condition leads to
\begin{equation}
 \Delta_p(v) -\Delta_d(v)= (j-k)\Delta\omega,
\label{eq:condition}
\end{equation}
with $\Delta\omega=2\pi/T_{\rm int}$. Using the definition for the Doppler shifted detunings, \refeqs{eq:DopplerDelta}, one obtains the velocity $v_{2ph}$ required for an atom to be transferred from \1 to \3:
\begin{equation}
 v_{2ph} = \frac{(j-k)\Delta\omega}{\omega_{13}}c,
\label{eq:v2ph}
\end{equation}
where we have assumed the nominal two-photon resonance condition $\Delta^{0}_{p} =\Delta^{0}_{d}$ and defined $\omega_{13}=\omega_{12}-\omega_{32}$.

One can obtain the same expression, \refeq{eq:v2ph}, through energy-conservation arguments as follows. In the reference frame of an atom moving with velocity $v$, each train of pulses in frequency domain corresponds to an optical frequency comb (OFC) with detunings 
\begin{subequations}
\label{eq:DetuningsDoppler}
\begin{align}
\Delta_{p}^{n}(v)&=\omega_{p}^{n}(v)-\omega_{12}, \\
\Delta_{d}^{m}(v)&=\omega_{d}^{m}(v)-\omega_{32}.
\end{align}
\end{subequations} 
Here the frequencies $\omega_{p}^{n}(v)=\omega_{p}^{0}(1+v/c)+n\Delta\omega$, and $\omega_{d}^{m}(v)=\omega_{d}^{0}(1+v/c)+ m\Delta\omega$ correspond to the different harmonics of the OFC, and $n$ and $m$ are the integer indices for each harmonic $(0, \pm 1,\pm 2,...)$, see Fig.~\ref{f:fig1}.
Therefore, transfer of atomic population from \1 to \3 will only be achieved by the simultaneous absorption of a pump photon and the stimulated emission of a dump one, such that their frequency difference matches the energy gained by the atom, \ie
\begin{equation}
	\Delta_{p}^{n}(v)=\Delta_{d}^{m}(v).
\label{eq:2phRes}
\end{equation}
This is the generalization of the two-photon resonance condition for pairs of harmonics of the two OFCs.  
It is easy to see that combining \refeqs{eq:DetuningsDoppler} with \refeq{eq:2phRes} and assuming $\Delta^{0}_{p} =\Delta^{0}_{d}$ we recover \refeq{eq:v2ph} by identifying $j-k=m-n$. Thus, every value of $v_{2ph}$, \ie every peak of the VC, is the result of the contribution of all the harmonics of both OFC fulfilling \refeq{eq:condition} and sharing the same value $j-k$.

Note that such a VC, governed by \refeq{eq:v2ph}, is not well defined for $\omega_{13}=0$, \ie for degenerate ground states. To physically understand this point, we should notice that in this situation the atoms, regardless of their velocity, see two OFCs that perfectly overlap in Fourier space since the field frequencies will have the same Doppler shift. Thus, a single possible match between harmonic indices occurs, $j=k$, which corresponds to $v_{2ph}=0$.

At the end of the PAP process, we expect to obtain a VC of population transferred to \3, $\rho_{33}(v)$, with peaks centered at velocity values given by \refeq{eq:v2ph}.

\begin{figure}[ht]
{
\includegraphics[width=1\columnwidth]{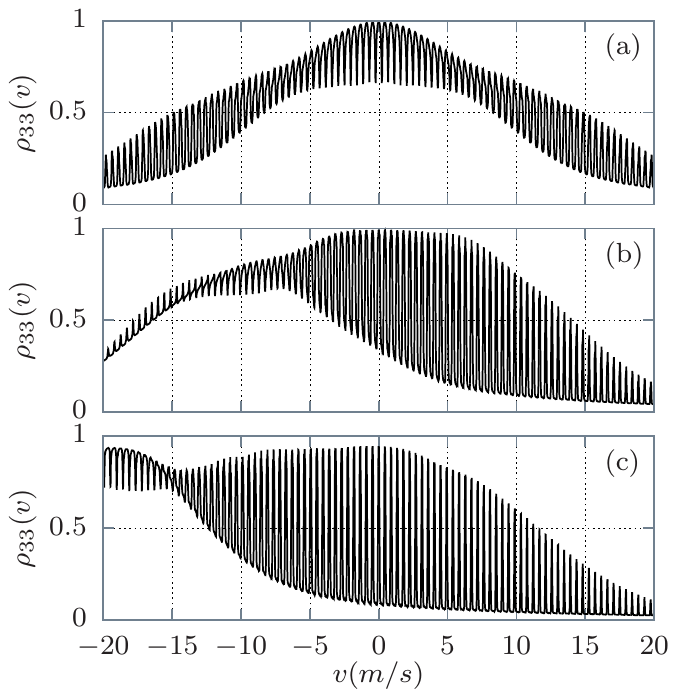}
}
\caption{Atomic VC via PAP generated using nominal detunings (a) $\Delta^{0}=0$, (b) $\Delta^{0}=2\pi\cdot63.7\,{\rm MHz}$, and (c) $\Delta^{0}=2\pi\cdot127.4\,{\rm MHz}$. Here $\rho_{33}$ is weighted by the Maxwell--Boltzmann velocity distribution with $\eta=350$ m/s and rescaled to show the single atom probability instead of the atomic density. See text for the rest of parameters.
}
\label{f:fig2}
\end{figure}

Finally, we have to take into account that for values of velocities different from those given in \eqref{eq:v2ph} the population can also be transferred to \3 by spontaneous emission after the absorption of one pump photon. This undesired effect will decrease the contrast of the velocity peaks around the central value $v=0$ distorting the comb.
We can diminish this effect by shifting the nominal detuning $\Delta^{0}(=\Delta_{p}^{0}=\Delta_{d}^{0})$ out of resonance.
Then, population decaying from \2 to \3 occurs for velocity classes far from $v=0$. A numerical example is shown in Fig.~\ref{f:fig2} for three different nominal detunings 
(a) $\Delta^{0}=0$, 
(b) $\Delta^{0}=2\pi\cdot63.7\,{\rm MHz}$, and 
(c) $\Delta^{0}=2\pi\cdot127.4\,{\rm MHz}$, using, in all cases, $N=18$ pump-dump pulses, $\omega_{32}=2\pi\cdot637$ THz, $\omega_{12}=2.5\,\omega_{32}$, $\Omega_{0}=2\pi\cdot80$ MHz, $\sigma=5$ ns, $T_{int}=0.7$ $\mu$s, and $\gamma_{21}=\gamma_{23}=10$ $\mu s^{-1}$.

\subsection{Atomic Frequency Comb (AFC)}
\label{sec:AtomicFrequencyCombAFC}

\subsubsection{AFC Configurations}

\begin{figure}[ht]
{
\includegraphics[width=0.8\columnwidth]{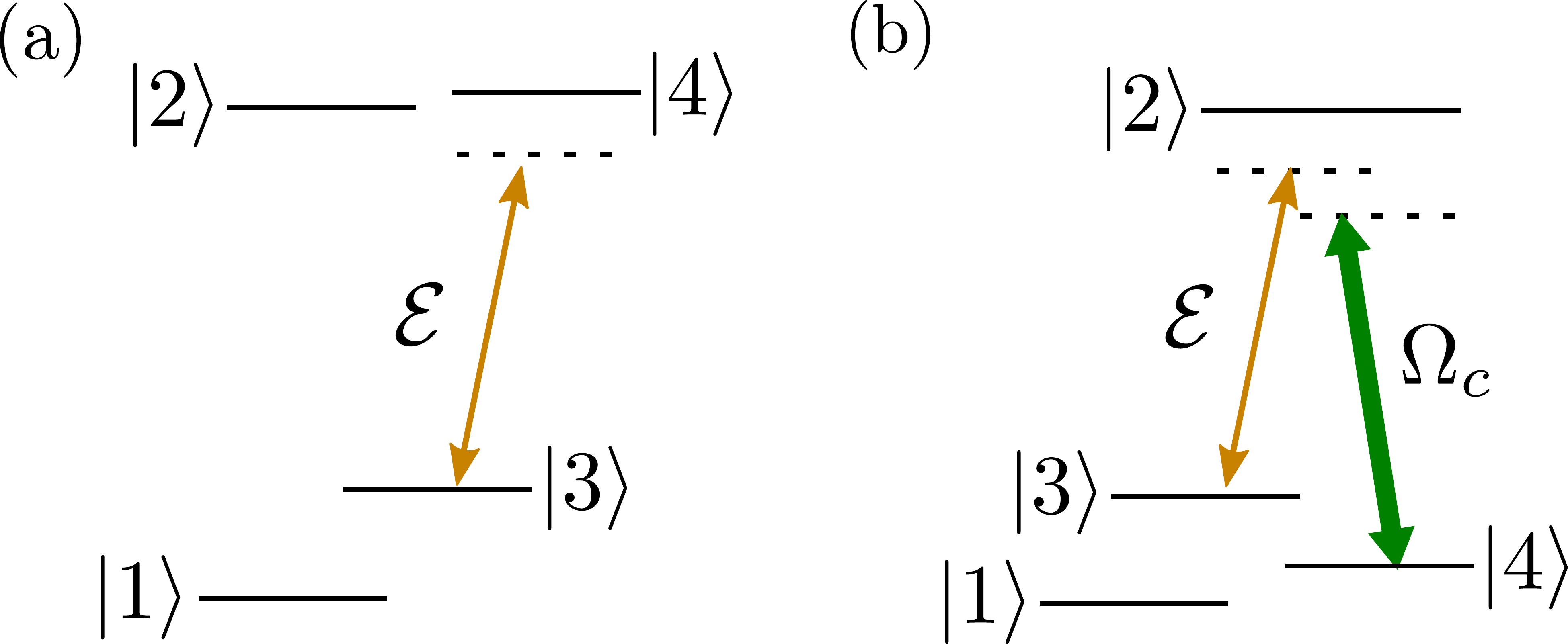}
}
\caption{Schematics of two AFC configurations to store a single photon. The signal photon $\mathcal{E}$ is stored in the \3$\leftrightarrow$\4 transition via a (a) one-photon or (b) two-photon process using a control field $\Omega_{c}$.}
\label{f:fig3}
\end{figure}

Once it is generated, the VC will translate into an absorption grating for a single photon pulse, $\mathcal{E}$, copropagating with the pump and dump laser pulses, coupling any transition involving state \3. Two different configurations for the storage process are shown in Fig.~\ref{f:fig3}. In the simplest configuration, one-photon AFC, a single photon directly couples the dipole allowed transition \3$\leftrightarrow$\4 [See Fig.~\ref{f:fig3}(a)], where \4 can be degenerate with state \2 or have any other energy. A second possibility is the two-photon based AFC where the single photon is mapped into the \3$\leftrightarrow$\4 transition via an off-resonant two-photon Raman process where, in this case, \4 is a metastable (long-lived) state [See Fig.~\ref{f:fig3}(b)].
The latter can allow, in principle, for longer memory lifetimes only limited by the finite width of the comb teeth and the atomic motion since the effect of the spontaneous emission from the excited states is avoided.
In the remaining of this section, we will focus only on the one-photon AFC configuration, although the results here obtained also apply for the two-photon configuration, provided that the state \2 can be adiabatically eliminated as we will discuss in Section \ref{sec:SIMULATION}.
We assume that, for atoms at rest, the signal photon is resonant with this transition, \ie $\delta^0\equiv\omega^0-\omega_{34}=0$, with $\omega^0$ being the central frequency of the photon wave-packet and $\omega_{34}$ the frequency of the transition \3$\leftrightarrow$\4. Therefore, for moving atoms, the corresponding detuning reads $\delta(v)=\omega^0v/c=\omega_{34}v/c$.

\subsubsection{AFC Parameters}

\begin{figure}[ht]
{
\includegraphics[width=1\columnwidth]{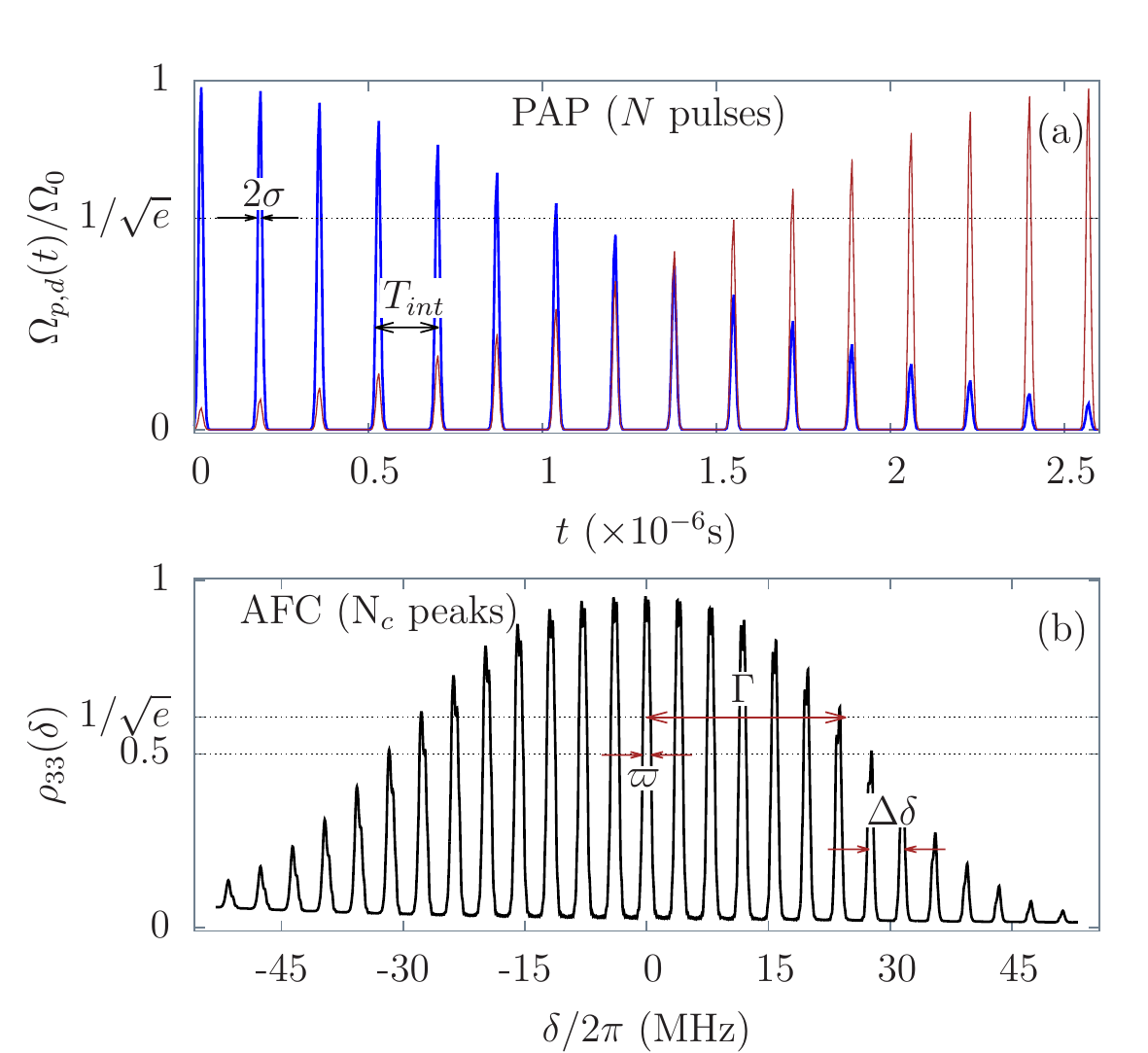}
}
\caption{Numerical example of the (a) temporal sequence of a piecewise adiabatic passage (PAP) with $N=16$ simultaneous dump (blue lines) and pump (red lines) pulses with temporal inter-pulse spacing $T_{int}$, and pulse width $\sigma$, which generates (b) an atomic frequency comb (AFC) with $\rm N_{c}$ peaks, bandwidth $\Gamma$, detuning peak separation $\Delta\delta$, and peak FWHM $\varpi$. Here $\eta=350$ m/s and $\rho_{33}(\delta)$ is rescaled to show the single atom probability instead of atomic density. See text for the parameter values.}
\label{f:fig4}
\end{figure}

In order to characterize the AFC, we model the density of atoms in state \3, after the transfer from \1 via PAP, with a sum of Gaussian functions, modulated by a Gaussian envelope of width $\Gamma$. Expressed as a function of the detuning $\delta$, it takes the form:
\begin{equation}
\label{eq:AFC}
\rho_{33}(\delta)= e^{-\delta^2/2\Gamma^2}\sum_{j=-\infty}^{\infty}e^{-(4\ln{2})(\delta-j\Delta\delta)^{2}/\varpi^{2}}.
\end{equation}
Here $\Delta\delta$ is the peak separation, while $\varpi$ is the FWHM of the AFC peaks.
A numerical example of the generation of the PAP train of pulses and of the corresponding AFC atomic spectral distribution is shown in Figs.~\ref{f:fig4}(a) and ~\ref{f:fig4}(b), respectively, with $N=16$ pump-dump pulses, $\omega_{34}=\omega_{32}=2\pi\cdot637$ THz, $\omega_{12}=2.5\,\omega_{32}$, $\Omega_{0}=2\pi\cdot151$ MHz, $\sigma=6.2$ ns, $T_{int}=0.17$ $\mu$s, $\Delta^{0}=2\pi\cdot 360$~MHz, and $\gamma_{21}=\gamma_{23}=10$ $\mu s^{-1}$.

Next, we discuss the figures of merit involved in the PAP-based AFC and derive their analytical expressions.

\paragraph{Bandwidth.}

The usual limitation for the comb width is given by the width of the Maxwell--Boltzmann distribution. However, here we consider only cases in which the comb width is narrower than the velocity standard deviation $\eta$, so the Doppler width does not affect the comb bandwidth.
To obtain an expression for the AFC bandwidth, we consider first the bandwidth of the respective OFCs for every train of pulses, which is given by the frequency-time Fourier relations, $1/\sigma$ (see Appendix A.1). Secondly, due to the Doppler effect, in terms of velocities, the pump and dump OFCs will have modified envelope bandwidths $c/\omega_{12}\sigma$ and $c/\omega_{32}\sigma$, respectively (see Appendix A.2). Moreover, the centers of these OFC depend on the atomic velocity. Thus, only those atoms with velocities producing a significant overlap between the two OFC will be able to interact with both fields. The overlap between the OFC envelopes (see Appendix A.2), in terms of the transition frequency $\omega_{34}$, has a bandwidth
\begin{align}
\label{eq:delta12}
\Gamma= &\frac{\sqrt{2}}{\sigma}\xi,
\end{align}
where $\xi\equiv\omega_{34}/\omega_{13}$.
For the AFC of the example in Fig. \ref{f:fig4}, the analytical (numerical) value is $\Gamma=2\pi\cdot24.19$ MHz ($2\pi\cdot25.46$ MHz).

Expression (\ref{eq:delta12}) gives us the bandwidth of the AFC with a Gaussian shape. We should mention that when increasing $\Omega_{0}$ or the number of pulses $N$, the adiabaticity of PAP increases and the height of the peaks in the comb rises up to its maximum value, given by the Maxwell--Boltzmann distribution (\ref{eq:MB}). When this occurs, the form of the comb acquires a flat top profile and expression (\ref{eq:delta12}) does not longer provide a good enough estimation of the comb bandwidth.

\paragraph{Peak separation.} Due to the condition given in \refeq{eq:v2ph} for the atoms to be transferred into state \3, the absorption peaks will be found at detunings fulfilling $\delta_{2ph}=\omega_{34}v_{2ph}/c$
and, consequently, they will be separated in frequency by
\begin{equation}
\label{eq:Deltadelta}
\Delta\delta=\xi\Delta\omega.
\end{equation}
Note that the $\xi$ factor in the last expression,
accounting for the asymmetry in the optical transition frequencies, determines the peak separation of the VC and, in turn, of the AFC.
For instance, if $\omega_{32}=\omega_{34}=\omega_{12}/3$, then $\Delta\delta=\Delta\omega/2$ which reduces by half the peak separation that one would have in a conventional AFC.
For the AFC of the example shown in Fig. 3, the analytical (numerical) value is $\Delta\delta=2\pi\cdot3.91$ MHz ($2\pi\cdot3.95$ MHz).

\paragraph{Number of peaks.} Considering $2\sqrt{2\pi}\Gamma$ for the base of the Gaussian comb, and for a large number of peaks, we can estimate the total number of peaks ${\rm N_c}$ forming the AFC with ${\rm N_{c}\Delta\delta}= 2\sqrt{2\pi}\Gamma$. Using \eqref{eq:delta12} and \eqref{eq:Deltadelta} we find that
\begin{eqnarray}
\label{eq:Nc}
	{\rm N_{c}}=\cfrac{2T_{int}}{\sqrt{\pi}\sigma},
\end{eqnarray}
which depends only on the PAP parameters. The ability to control the number of peaks of the AFC is an interesting feature since this quantity determines the number of temporal modes that can be stored when used as a quantum memory \cite{Afzelius'09}. For the AFC of the example of Fig. 4, the analytical (numerical) value is $\rm{N_{c}}=25.7$ peaks ($25$ peaks).

\paragraph{Peak width.}
The FWHM of each individual tooth of the AFC is given by the transfer efficiency under conditions of quasi-two-photon resonance. Based on the expression for the velocity peak width obtained in STIRAP for a Doppler broadened medium and the relationship between the PAP and STIRAP processes (see Appendix B), one can estimate the width of each peak as
\begin{align}
\label{eq:FWHMFrequency}
\varpi=\frac{\sqrt{\pi}\Omega_{0}^2\sigma\xi}{4\Delta^{0}T_{int}},
\end{align}
for $|\Delta^{0}|>\Omega_{0}\sqrt{\omega_{32}/\omega_{12}}$.
For the AFC of the example in Fig. \ref{f:fig4}, the analytical and numerical values are $\varpi=2\pi\cdot0.684$ MHz and $\varpi=2\pi\cdot0.891$ MHz, respectively. The discrepancy is due to the fact that expression (\ref{eq:FWHMFrequency}) has been derived neglecting spontaneous emission, in the limit of large detunings, for combs with a large number of peaks, and assuming perfect Gaussian profiles. 

\paragraph{Finesse.} One of the important parameters to estimate the quality of an AFC is the finesse of the comb $\mathcal{F}\equiv\Delta\delta/\varpi$. In terms of the system parameters, it can be written as
\begin{align}
\label{eq:finesse}
\mathcal{F}=\frac{8\sqrt{\pi}\Delta^{0}}{\Omega_{0}^2\sigma}.
\end{align}
According to this expression, the finesse of the comb does not depend on the number of pulses but can be varied adjusting the rest of the PAP parameters ($\Omega_{0},\sigma,\Delta^{0}$). For the AFC of the example in Fig. \ref{f:fig4}, the analytical and numerical values are $\mathcal{F}=5.72$ and $\mathcal{F}=4.43$ ,respectively.
The discrepancy comes from the assumptions done in the estimation of the peak width $\varpi$, commented above.


In what follows, we summarize the differences of the AFC created via PAP with respect to the conventional AFC generated by a train of pulses with the same pulse temporal spacing:

(i) The form of the frequency peaks is not affected by the temporal shape of the individual pulses. This is because our comb is generated via PAP, whose efficiency is insensitive to the shape of the pulses, provided that their envelopes follow the counterintuitive sequence of STIRAP.

(ii) Performing PAP more adiabatically, e.g., by increasing $\Omega_{0}$ or $N$, produces an increment of the height of the peaks and thus enhances the optical depth of the $\3\leftrightarrow\4$ transition.

(iii) The peaks frequency spacing, $\Delta\delta$, is proportional to $2\pi/T_{int}$ but also to the ratio of the transition frequencies of the $\Lambda$-type system and the storage transition, via the $\xi$ factor.

\subsection{AFC-based QM}
\label{sec:QuantumMemoryQM}

From \eqref{eq:Deltadelta}, one can derive the expression for the retrieval time $\widetilde{\mathcal{T}}$ 
in an AFC-based QM via PAP as
\begin{eqnarray}
\label{eq:TN1}
\widetilde{\mathcal{T}}&=&\cfrac{T_{int}}{\xi}=\cfrac{\omega_{13}}{\omega_{34}}\,T_{int},
\end{eqnarray}
%
where $T_{int}=2\pi/\Delta\omega$ coincides with the retrieval time of the conventional AFC-based QM. As previously commented, \refeq{eq:TN1} implies that, when our AFC proposal is used as a quantum memory, the retrieval time also depends on the chosen specific atomic system through $\xi$.

Next, we discuss the parameters for PAP and the frequencies of the transitions involved in order to optimize the QM and satisfy the assumptions required in our model.

On the one hand, in order to achieve high storage efficiencies
a large optical depth is needed. In turn, the higher the finesse the better the reemission due to the atoms rephasing. However, it is easy to show that the optical depth of the medium is effectively reduced by a factor $1/\mathcal{F}$. Therefore, a compromise between a large optical depth and the reemission efficiency is required. It has been shown that for a backward retrieval scheme, in the absence of spontaneous emission, the memory efficiency reads $\eta\simeq(1-e^{-OD/\mathcal{F}})^2e^{-7/\mathcal{F}}$ \cite{Afzelius'09}, where $OD$ is the optical depth of the medium. Thus, for large optical depths, when the efficiency is above 90\% for $\mathcal{F}\gtrsim10$.
Using \refeq{eq:finesse} with $\Gamma\gg\Delta\delta$, this condition for the finesse leads to
\begin{align}
\label{eq:PAPoptimal}
\frac{\Omega^{2}_{0}\sigma}{\Delta^{0}}\lesssim\frac{4}{5}\sqrt{\pi}\approx1.4.
\end{align}
%
From this expression it is clear that in order to obtain a large finesse, for a given $\Omega_{0}$ and $\sigma$, we must increase the nominal detuning. This is even more clear looking at Fig.~\ref{f:fig2} where the widths of the velocity peaks decrease for increasing detunings. 
Moreover, to avoid spontaneous emission for the central velocity classes, a  necessary requirement would be to impose a nominal detuning larger than the comb bandwidth itself, given in \refeq{eq:delta12}. 

On the other hand, if we define the ratio between the transition frequencies of the $\Lambda$-system as $r\equiv\omega_{12}/\omega_{32}$, so such that $\xi=(\omega_{34}/\omega_{32})/(r-1)$, one can see that our AFC is not formally available for degenerate ground states, \ie $r=1$, since according to \eqref{eq:Deltadelta}, $\Delta\delta\rightarrow\infty$, which it is consistent with the discussion about the velocities given in \eqref{eq:v2ph} in \ref{sec:VelocityCombVC} when $\omega_{13}=0$.

Furthermore, according to \refeq{eq:TN1}, to obtain longer retrieval times than in conventional AFC-based QMs, it should be satisfied $\xi<1$, that is,
\begin{equation}
\label{eq:ratio}
\omega_{34}<\omega_{13},
\end{equation}
or equivalently, $r>\omega_{34}/\omega_{32}+1$, which becomes simply $r>2$ if \2 and \4 are degenerate.

To summarize, \refeq{eq:PAPoptimal} and \refeq{eq:ratio} give us the 
conditions to implement an AFC-based QM via PAP, which improve some features of conventional AFC-based QMs.

\section{NUMERICAL EXAMPLE OF A PAP-based AFC-QM}
\label{sec:SIMULATION}

In this section, we show a numerical example of the proposed AFC-based QM via PAP. We consider Ba atoms \cite{Jitschin'80}, although other atomic species with different energy-level schemes could be chosen. The reasons for this choice may be summarized as follows: (i) Alkaline earth metals such as calcium or barium feature available $\Lambda$-type systems with a large asymmetry in the pump and Stokes transitions, formed by $S$, $P$, and $D$ states, where the later are, very often, long-lived metastable states. This configuration allows us to fulfill Eq.~\eqref{eq:ratio}; (ii) They also have transitions in the telecommunication range ($\sim1.5\,\mu$m), which are of special interest for long distance quantum communications \cite{Rielander'16,Rancic'17}; and (iii) Alkaline earth metals can be prepared in vapor cells or hollow cathode lamps capable of reaching high atomic densities \cite{Fang'17}. 

\begin{figure}[ht]
{
\includegraphics[width=0.8\columnwidth]{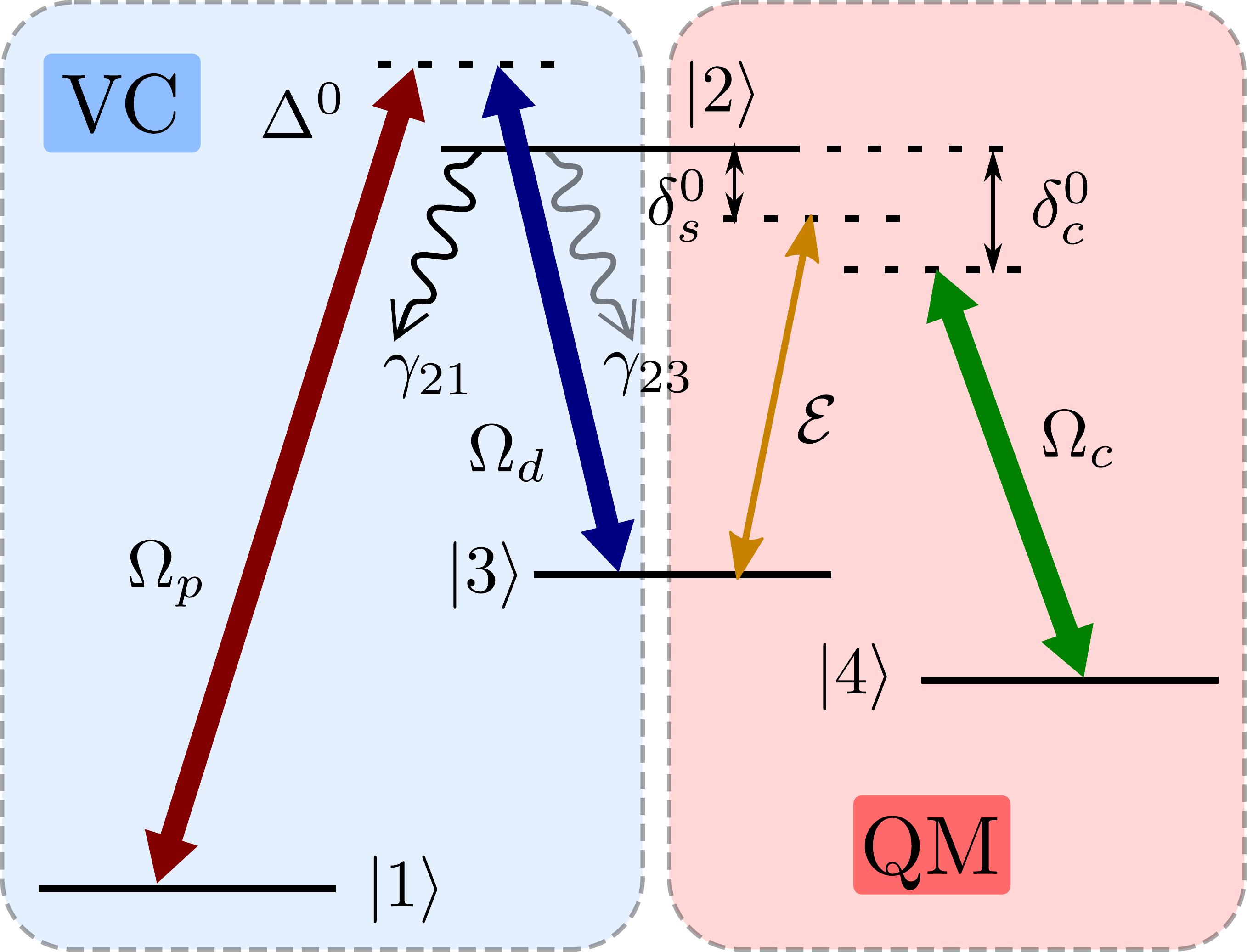}
}
\caption{Relevant energy-levels for Ba with the Rabi frequencies and detunings involved in the VC creation via PAP (left frame) and the single telecom photon QM (right frame). See the main text for the definition of the parameters.
}
\label{f:fig5}
\end{figure}

For the simulation of the VC creation we consider the $\1=6s^2\,{^1}S_0\leftrightarrow\2=6s6p\,{^1}P_1$ and $\2\leftrightarrow\3=6s5d\,{^1}D_2$ transitions of Ba, see Fig.~\ref{f:fig5} (left frame). Since in this system the spontaneous emission from the excited state \2 occurs predominantly towards the initial ground state \1, the number of atoms that can reach the final state \3 via spontaneous decay is very limited. Moreover, the $\2\leftrightarrow\3$ transition is already in the telecom range (1500.4 nm). For this reason, to implement the QM (once the VC has been created), one could consider a single telecom photon coupling the $\2\leftrightarrow\3$ transition. However, in order to circumvent the spontaneous emission from the excited state $\2$, the photon will be stored in the atomic coherence between state \3 and the long-lived state $\4=6s5d\,{^3}D_2$ through an off-resonant two-photon Raman process [Fig.~\ref{f:fig5} (right frame)].


First, for the simulation of the AFC preparation we use the density matrix equations and parameters for a Ba vapor (see transition properties in, e.g., \cite{DammalapatiThesis}) at 800~ºC (typical in hollow cathode lamps \cite{Dammalapati'09,Araujo'08,Fang'17}) and a density of $\varrho=2.5\times10^{20}$~at/m$^3$. The decay rates from the excited state to the ground and metastable states are $\gamma_{21} = 1.19\times10^{8}$ s$^{-1}$ and $\gamma_{23} = 0.25\times10^{6}$ s$^{-1}$ (the decay to state $\4$ is $\gamma_{24}\simeq\gamma_{23}/100$ so we can safely neglect it), respectively, while the transition frequencies are $\omega_{32}=2\pi\cdot200$ THz and $\omega_{12}=2\pi\cdot540$ THz, so $\omega_{13}=2\pi\cdot340$ THz.
The effective storage transition $\3\leftrightarrow\4$ has a frequency $\omega_{34}\equiv\omega_{42}-\omega_{32}=2\pi\cdot65.35$ THz, where $\omega_{42}=2\pi\cdot270$ THz (1130.6 nm). With these values we obtain $\xi=0.19$ and $r=2.7$, satisfying condition \eqref{eq:ratio}. 
We perform numerical simulations using different number of pulses, Rabi frequencies, and nominal detunings. In all cases the pulses of the train are separated by $T_{int}=689$~ns and have a duration $\sigma=T_{int}/64=10.77$~ns. The envelope duration, $\sigma_{e}=\tau/(2\sqrt{2\ln2})$, is chosen depending on the number of pulses and the nominal detunings are such that the two-photon resonance condition is satisfied for the central velocity class $\Delta^{0}(=\Delta_{p}^{0}=\Delta_{d}^{0})$. 
For these parameters, the estimated peak separation is $\Delta\delta=2\pi\cdot0.28$~MHz, corresponding to a retrieval time of $\widetilde{\mathcal{T}}=3.6\,\mu$s. 

After the comb creation, we simulate the propagation of a signal photon 
coupling transition $\2\leftrightarrow\3$, see Fig.~\ref{f:fig5}, through a medium of length $L=2$ cm.
We consider that at the input of the medium, the photon 
wavepacket can be described with a slowly varying Gaussian envelope
\begin{equation}
\label{Eini}
\mathcal{E}(z=0,t)=\cfrac{1}{\sqrt{\tau_{p}\sqrt{\pi}}}\,e^{-(t-t_{c})^2/(2\tau_{p}^2)},
\end{equation}
normalized such that $\int{\left|\mathcal{E}(0,t)\right|^2dt=1}$. Here, $\tau_{p}=0.3\,\mu$s is the duration of the pulse, centered at $t_{c}=4\tau_{p}$. Since the excited state \2 has a short lifetime, we consider the signal photon being tuned out of resonance, with detuning $\delta_{s}^{0}$, and it is coupled via a strong control field, of Rabi frequency $\Omega_{c}$, to the $\4$ metastable state in a two-photon Raman configuration. In this configuration, the harmful effect of the spontaneous emission is reduced and the signal photon is effectively stored in the $\3\leftrightarrow\4$ atomic coherence.
Thus, by adiabatically eliminating the excited state, the equations describing the propagation of the signal photon and the slowly-varying spin wave amplitudes, $\mathcal{E}$ and $\mathcal{S}$, respectively, in the photon co-moving frame $t\rightarrow t-z/c$ \cite{Gorshkov'07} read
\begin{align}
	\partial_{z}\mathcal{E}(z,t)=&-i\frac{g^2\varrho}{c}\int{\frac{\rho_{33}(v)}{\delta_{s}(v)+i\gamma}d\delta}\mathcal{E}(z,t) \nonumber \\
    &-i\frac{g\sqrt{\varrho}}{c}\Omega_{c}\int{\frac{\rho_{33}(v)}{\delta_{s}(v)+i\gamma}\mathcal{S}(z,t,\delta)d\delta}, \label{eq:PhotProp} \\
    \partial_{t}\mathcal{S}(z,t,\delta)=&i\left\{\delta(v)-i{\rm Im}\left[\frac{\Omega_{c}^2}{\delta_{s}(v)+i\gamma}\right]\right\}\mathcal{S}(z,t,\delta) \nonumber \\
    &-i\frac{g\sqrt{\varrho}\Omega_{c}}{\delta_{s}(v)+i\gamma}\mathcal{E}(z,t). \label{eq:SpinEvol}
\end{align}
Here $c$ is the speed of light in vacuum, $g=\mu_{23}\sqrt{\frac{\omega_{32}}{2\epsilon_{0}\hbar}}$ the photon-atom coupling constant, with $\epsilon_{0}$ the vacuum electric permittivity, $\gamma$ is the coherence decay rate (which we assume $\gamma\simeq\gamma_{21}/2$, since $\gamma_{23}$ is negligible), and $\delta(v)\equiv\delta_{s}(v)-\delta_{c}(v)-{\rm Re}\{\Omega_{c}^2/[\delta_{s}(v)+i\gamma]\}$ the effective two-photon detuning, with $\delta_{s}(v)=\delta_{s}^{0}+\frac{v}{c}\omega_{32}$ and $\delta_{c}(v)=\delta^{0}_{c}+\frac{v}{c}\omega_{42}$, the signal and control detunings, respectively, including the Doppler shifts.
In order for the adiabatic elimination of state \2 to hold, we must impose a large detuning compared to the other frequency scales of the system, thus we choose $\delta_{s}^{0}\simeq-2\pi\cdot380.38$ MHz and $\Omega_{c}=2\pi\cdot15.20$ MHz. The control detuning is set to be on two--photon resonance with the probe detuning and to compensate the AC-Stark shift $\delta^{0}_{c}=\delta_{s}^{0}-\Omega_{c}^2/\delta_{s}^{0}$. With this choice of parameters the effective detuning in Eq.~(\ref{eq:SpinEvol}) is proportional to the frequency difference between the signal and control transitions $\delta(v)\simeq\omega_{34}v/c$, and the atoms are on quasi two-photon resonance with the fields when $v=0$. For convenience, we perform a backward retrieval protocol \cite{Afzelius'09} for which the retrieved field is always maximum at $z=0$.


\begin{figure}[ht]
{
\includegraphics[width=1\columnwidth]{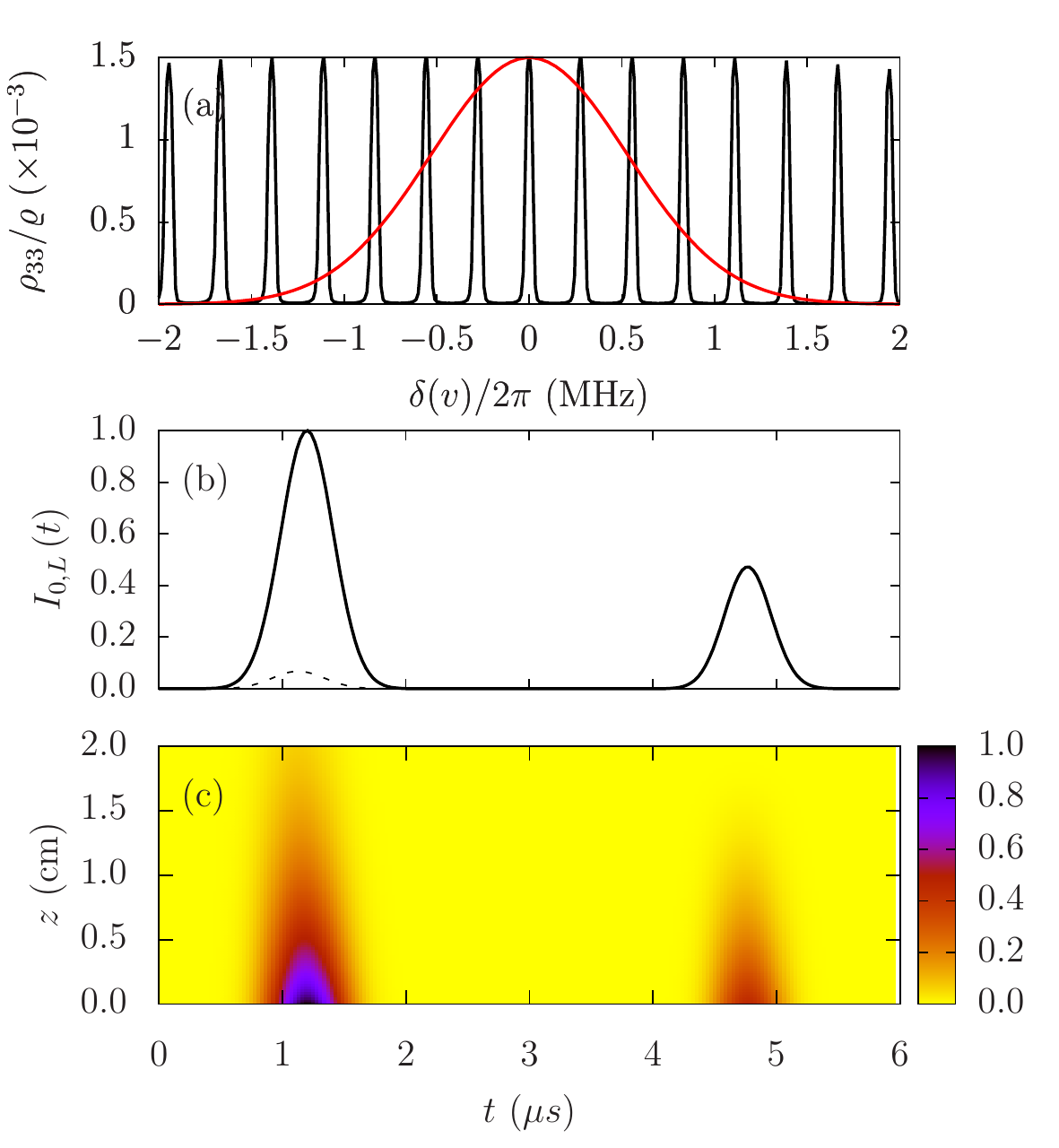}
}
\caption{(a) Comb distribution $\rho_{33}/\varrho$ (black line) as a function of the detuning $\delta(v)$, together with the input photon spectral distribution seen by the atoms (red line). (b) Temporal evolution of the photon intensity at $z=0$ (solid line) and $z=L$ (dashed). The retrieved pulse in the backward direction is found at $\widetilde{\mathcal{T}}\simeq3.6\,\mu$s respect to the input pulse. (c)  Scaled photon intensity as a function of position and time. The simulation is performed with an AFC comb created using $N=40$ pulses, $\Omega_0=2\pi\cdot51.72$ MHz, and $\Delta^{0}\simeq2\pi\cdot129.35$ MHz (see text for the rest of parameters).
}
\label{f:fig6}
\end{figure}


The simulation is performed by discretizing the space and time variables, and considering a large enough number of velocity classes to ensure the convergence of the results. The integration time is $t_f=2t_c+\widetilde{\mathcal{T}}$. For the simulation we choose only a section of the comb, but still larger than the incident photon bandwidth, see Fig.~\ref{f:fig6}(a). For this figure, the parameters used to create the comb are $N=40$ pulses, $\Omega_0=2\pi\cdot51.72$ MHz, and $\Delta^{0}=8.75/\sigma=2\pi\cdot129.35$ MHz,  resulting in a comb finesse of $\mathcal{F}=5.9$ (the peak separation is $\Delta\delta
=2\pi\cdot0.28$ MHz and the peak width $\varpi=2\pi\cdot47.27$ kHz).

In Figs.~\ref{f:fig6}(b,c) we plot the scaled intensity, $I(z,t)\equiv\left|\mathcal{E}(z,t)\right|^2/\left|\mathcal{E}(0,t_{c})\right|^2$, of the signal photon. In Fig.~\ref{f:fig6}(b) the solid line corresponds to $I_{0}(t)\equiv I(0,t)$, where the left and right Gaussian profiles indicate the intensity of the incident and backward-retrieved photons, respectively. Moreover we show the transmitted intensity of the incident photon $I_{L}(t)\equiv I(L,t)$ with a dashed line. In Fig.~\ref{f:fig6}(c) we show the photon scaled intensity during the full propagation, as a function of time and space. 
For this example, we obtain that the incoming photon is absorbed into the medium with a storage efficiency of $\eta_{s}\equiv 1-\int^{\dv{t_f/2}}_{0}{\left|\mathcal{E}(L,t)\right|^2dt}=93.4\%$, and that the reemission time occurs around $\widetilde{\mathcal{T}}\equiv T_{int}\omega_{13}/\omega_{34}=3.6\,\mu$s respect to the input pulse, as expected from \refeq{eq:TN1}
. 
The retrieval efficiency is $\eta_{r}\equiv\int^{t_f}_{t_f/2}{\left|\mathcal{E}(0,t)\right|^2dt}=41.3\%$ (typical experimentally reported AFC efficiencies are around 15-20\% \cite{Rielander'14}), limited partially due to imperfect comb profile and a low effective optical depth,
which causes part of the photon wavepacket to leak at the medium output.
\begin{figure}[t!]
\centering
\includegraphics[width=1\columnwidth]{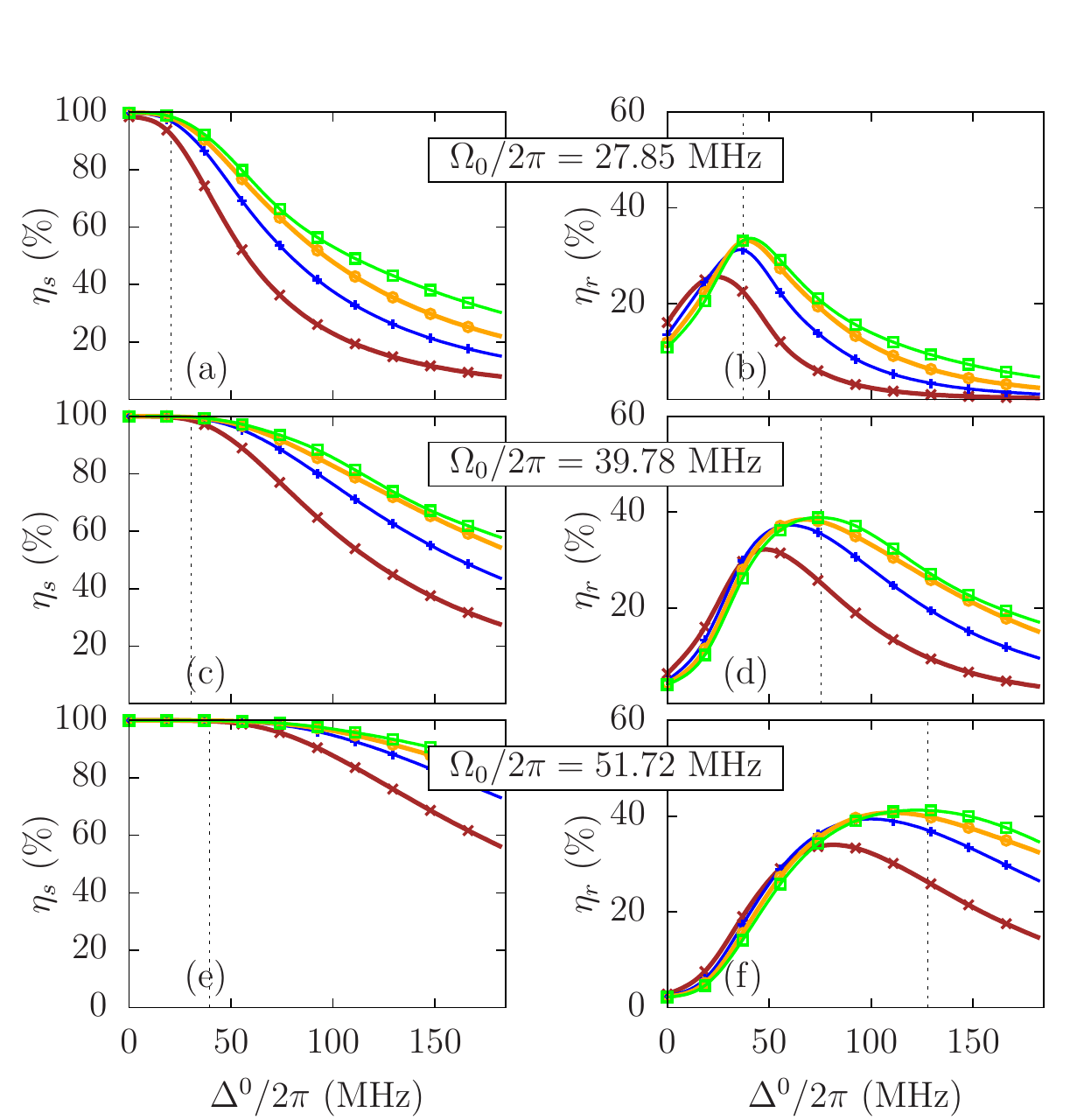}
\caption{(a,c,e) Storage and (b,d,f) retrieval efficiency for the signal photon as a function of the nominal detuning used to produce the AFC for pulse number $N=10$ (red lines-crosses), $N=20$ (blue lines-plus signs), $N=30$ (yellow lines-circles), and $N=40$ (green lines-squares), and Rabi frequency (a,b) $\Omega_{0}=2\pi\cdot27.85$ MHz, (c,d) $\Omega_{0}=2\pi\cdot39.78$ MHz, and (e,f) $\Omega_{0}=2\pi\cdot51.72$ MHz.  The rest of parameters are as in Fig.~\ref{f:fig5}. The vertical dashed lines correspond to the values given by \refeq{eq:umbral} (a,c,e) and \refeq{eq:PAPoptimal} (b,d,f).}
\label{f:fig7}
\end{figure}


In Fig.~\ref{f:fig7} we show the storage (a,c,e), and retrieval (b,d,f) efficiencies as a function of the nominal detuning imposed during the AFC creation, for different number of pulses $N=10$ (red lines-crosses), $N=20$ (blue lines-plus signs), $N=30$ (yellow lines-circles), and $N=40$ (green lines-squares), and three Rabi frequencies for the pump and dump fields (a,b) $\Omega_{0}=2\pi\cdot27.85$ MHz, (c,d) $\Omega_{0}=2\pi\cdot39.78$ MHz, and (e,f) $\Omega_{0}=2\pi\cdot51.72$ MHz.

In the first column [Figs.~\ref{f:fig7} (a,c,e)], the storage efficiency $\eta_{s}$ slightly changes for small values of $\Delta^{0}$ and then experiences a pronounced decrease for increasing values of the detuning.
This behavior can be understood by considering that there is a value of the detuning given by \refeq{eq:umbral} in Appendix B, \ie 
$\Delta^{0}/\Omega_{0}=\sqrt{\omega_{32}/\omega_{13}}$ (indicated by vertical dashed lines), above which the width of the absorption peak changes from being approximately constant to asymptotically decrease with $\Delta^{0}/\Omega_{0}$.

In the second column [Figs.~\ref{f:fig7} (b,d,f)] we observe maxima values for the  retrieval efficiency $\eta_{r}$.
As demonstrated in \cite{Afzelius'09}, there is an optimal finesse ($\gtrsim10$ for large optical depths) which maximizes the retrieval efficiency.
This is consistent with the observed maxima, which coincide approximately with the value given by condition (\ref{eq:PAPoptimal}), represented as vertical dashed lines. We observe that, for large detunings, a large number of pulses (or Rabi frequency) improves the efficiency since it increases the height of the comb peaks (and hence the optical depth). On the contrary, in the region of small values of the detuning, we observe that for a low number of pulses or small Rabi frequencies, the efficiency is slightly higher. This is because in this region the population transfer via spontaneous decay is more significant. Therefore, the larger the number of pulses or the Rabi frequency the larger the amount of undesired population is transferred to \3 from state \2. 

It is worth to mention that even with the mentioned limitations, the large atomic densities achievable in hot vapors provide, in general, large efficiencies compared to, \eg cold atoms QMs. In particular, for the parameters used in our example, the efficiencies are larger than the state-of-the-art AFCs in REICs. Moreover, the number of pulses used to achieve such efficiencies is remarkably lower than for conventional AFC creation methods. 
Nevertheless, we must point out here that the efficiencies depend strongly on the Rabi frequencies used. For instance, the transition $\2\leftrightarrow\4$ exhibits a weak dipole moment, which would require large intensities for the driving laser. In particular, the value of the Rabi frequency $\Omega_{c}$ considered in  Fig.~\ref{f:fig6} corresponds to an intensity of $\sim2.4$ W/cm$^2$. This requirement could be reduced by, \eg tailoring the temporal shape of the $\Omega_{c}$ to have a better matching with the single photon, but this lies out of the scope of this work.
We should also note that the retrieval time could be partially controlled by switching off the control field during the storage interval. Differently from the typical spinwave AFC \cite{Gundogan'15,Timoney'12,Yang'18}, in our case the evolution of the coherence for each atom does not stop during the storage of the photon, but it evolves with a phase depending on the atom velocity due to the Doppler shift $v\omega_{34}/c$.
Therefore, at the moment of rephasing, if the control field is switched off the atoms cannot reemit the photon. Thus, one could wait until a subsequent rephasing to turn on the control field and retrieve the photon at a time multiple of the original retrieval time (2nd, 3rd, 4th... echo). In any case, since state \4 is metastable this would allow to store the photon for times only limited by atomic motion. Note that the latter has not been considered in this work since the numerical simulations presented here do not aim at reproducing a realistic experiment but at guiding experimentalists in the basic steps to implement our protocol. In any case, however, an increasing number of experiments are now focused on  implementing QMs in hot vapors \cite{Reim'10,Reim'11,Hosseini'11,Finkelstein'18,Kaczmarek'18,Guo'18} in which the detrimental effects of atomic motion are avoided, e.g., by using buffer gases, cell coatings, specific Raman configurations, and short enough signal photons.

\section{CONCLUSIONS}
\label{sec:CONCLUSIONS}

In this work, we have first studied the implementation of a VC and, eventually, an AFC in hot atomic vapors by using the PAP technique to, later one, discuss its application for QMs. We have shown that by using this technique a reduced number of pulses, compared to standard methods, is enough to create a well defined AFC, whose properties not only depend on the applied trains of pulses, but also on the specific atomic transition frequencies considered.
In particular, the peak separation of the comb can be several times smaller than the one that would be obtained with conventional excitation techniques. Moreover, due to the adiabatic following of a dark-state involved in the PAP technique, the resulting combs are robust under intensity fluctuations of the pulses. We have derived analytical expressions for the comb characteristic parameters such as the bandwidth, peak separation, number of peaks, peak width and finesse, and determined the optimal conditions for its application in QMs.  In particular, we have studied the implementation of this technique in a high density Ba atomic vapor for the storage and retrieval of single photons at the telecom range.
Finally, although this technique has been discussed for hot vapors in which the Doppler effect is exploited, it could also be implemented in other $\Lambda$-type systems with a large enough inhomogeneous broadening in the two-photon transition due to, e.g., inhomogeneous magnetic fields. 

\section*{ACKNOWLEDGMENTS}
\label{sec:ACKNOWLEDGMENTS}

The authors gratefully acknowledge financial support
through the Ministerio de Economía y Competitividad (MINECO) (FIS2014-57460-P, FIS2017-86530-P) and from the Generalitat de Catalunya (SGR2017-1646).

\section*{Appendix A}
\label{sec:AppendixA}

\subsection*{\textbf{A.1} OFC bandwidth and height}
\label{sec:AppendixA1}

We consider a train of $N$ decreasing pulses as the one in \refeq{eq:fieldD}. If we assume that the individual pulses, of width $\sigma$, are much shorter than the Gaussian envelope, of width $\sigma_{e}$, the train can be approximated to
\begin{equation}
\label{eq:ApendixA1}
\Omega(t)=\Omega_{0}\sum_{n=0}^{N-1}\Omega_{n}\,e^{-(t-n{T_{int}})^{2}/2\sigma^2},
\end{equation}
where $\Omega_{n}
=e^{-n^2T_{int}^2/2\sigma_{e}^{2}}$ being $T_{int}$ the separation between two consecutive pulses.

The corresponding OFC is given by the absolute value of the Fourier transform of \refeq{eq:ApendixA1}, defined as $\tilde{\Omega}(\omega)\equiv\int_{-\infty}^{+\infty}\Omega(t)e^{-i\omega t}d\omega$, which reads
\begin{equation}
\label{eq:ApendixA2}
\tilde{\Omega}(\omega)=\sigma\Omega_{0}e^{-\omega^2\sigma^2/2}\sum_{n=0}^{N-1}\Omega_{n}e^{inT_{int}\omega},
\end{equation}
where the envelope has a bandwidth of $1/\sigma$.
The maxima of the OFC peaks correspond to the frequencies $\omega=k2\pi/T_{int}$, with $k\in\mathds{Z}$, since for those values the summation of \refeq{eq:ApendixA2} takes its maximum value, $\sum_{n=0}^{N-1}\Omega_{n}$, which increases with $N$. Thus, the height of the OFC increases with $N$ and it is proportional to $\Omega_{0}$.


The same results can be obtained for a train of $N$ increasing pulses like the one given by \refeq{eq:fieldP}.



\subsection*{\textbf{A.2} AFC bandwidth}
\label{sec:AppendixA2}

To obtain an expression for the bandwidth of the AFC, we proceed as follows.
From \eqref{eq:ApendixA2}, the envelopes of the pump and dump OFCs for an atom at rest are two Gaussians whose centers are given by the sum of the corresponding transition frequency and nominal detuning, 
\begin{align}
	S_{p,d}(\omega)\propto\exp{\left[-\frac{\left(\omega-(\omega_{12,32}+\Delta_{p,d}^{0})\right)^2}{2/\sigma^2}\right]},
\end{align}
where $\sigma$ is the width of the field pulses.
Further, for a moving atom, the Doppler shift changes the detuning according to \refeq{eq:DopplerDelta} so, the previous expression reads 
\begin{align}
	S_{p,d}(\omega)\propto\exp{\left[-\cfrac{\left(v-c\,\cfrac{\omega-\omega_{12,32}-\Delta_{p,d}^{0}}{\omega_{p,d}^{0}}\right)^2}{2\left(\cfrac{c}{\sigma\omega_{p,d}^{0}}\right)^2}\right]},
\end{align}
whose respective bandwidth in velocity is $c/(\sigma\omega_{p,d})^{0}$.
We note that each spectral distribution is displaced differently depending on the atomic velocity.
Thus, only the atoms that experience a non-zero overlap with the two fields $|\int{S_{p}(\omega)S_{d}(\omega)}d\omega|$ will have the possibility to be transferred to state \3 via PAP. It is easy to see that such an overlap takes the form of a Gaussian with a width $\sqrt{2}c/(\sigma\omega_{13})$, which determines the range of velocities that allows the atoms to interact with both fields. Finally, in terms of the storage transition frequency $\omega_{34}$, we recover the bandwidth of the AFC in \refeq{eq:delta12}, $\Gamma=\sqrt{2}\omega_{34}/(\sigma\omega_{13})=\sqrt{2}\xi/\sigma$. 

\section*{Appendix B}
\label{sec:AppendixB}

\begin{figure*}[ht]
{
\includegraphics[width=1\textwidth]{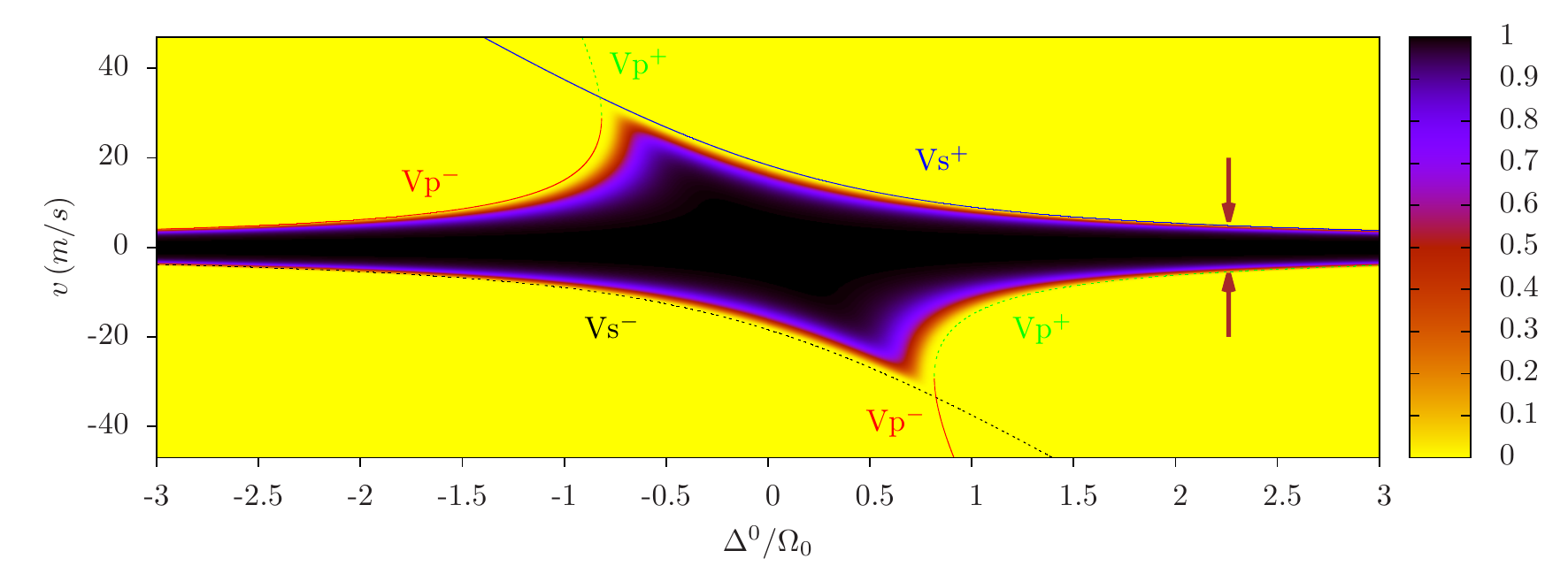}
}
\caption{Density plot of $\rho_{33}$ via STIRAP as a function of the atomic velocity $v$ and the ratio $\Delta^{0}/\Omega_{0}$. Note that the region in which the population is transfered to \3, named \textit{Optimal Zone}, is bounded by the curves ${\rm V_{s,p}^{\pm}}$. The values for the frequencies, couplings, and decays are the same as in Fig.~\ref{f:fig4}.
}
\label{f:fig8}
\end{figure*}

To obtain the expression for the width of the peaks of the AFC, i.e.,  \refeq{eq:FWHMFrequency}, we proceed in two steps as follows.

\begin{figure}[ht]
{
\includegraphics[width=1\columnwidth]{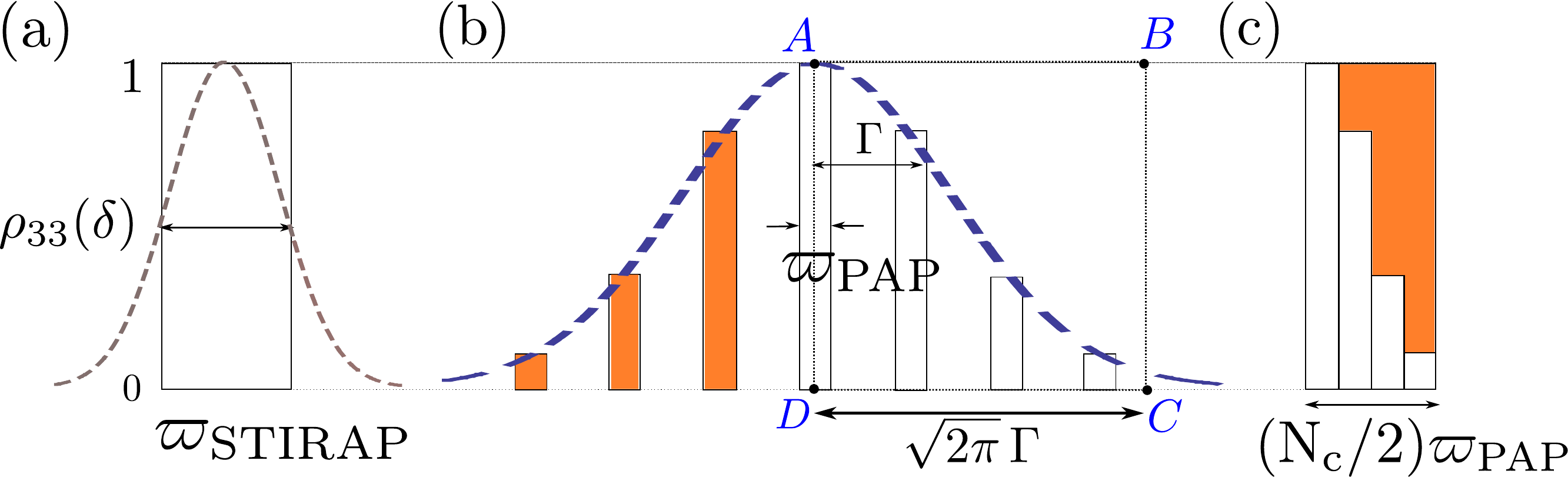}
}
\caption{Modeling of the $\rho_{33}(\delta)$ profile for (a) STIRAP and (b) PAP with rectangles of the same area. We can reconstruct the rectangle of (a) by combining conveniently the ${\rm N_{c}}$ peaks of the comb (b) as shown in (c). This allows us to relate the widths of the STIRAP and PAP peaks (see text).
}
\label{f:fig9}
\end{figure}

\textit{I. Frequency peak for STIRAP}. 
In STIRAP, a $\Lambda$-system as the one shown in Fig.~\ref{f:fig1} (and following the same notation and definitions for clarity) is shined by a pair of pulses, pump and dump, which are sent in a counterintuitive temporal manner. 
In order for STIRAP to properly work, the pump and dump fields
have to fulfill the so called two-photon resonance condition, i.e., $\Delta^0_{p}-\Delta^0_{d}=0$. Beyond this condition,
it can be shown that the final population $\rho_ {33}$ after applying STIRAP is distributed in a two-photon resonance window \cite{Vitanov'10}. Similarly, for a Doppler broadened medium the velocity of an atom $v$ produces a differential Doppler shift in the pump and dump transitions, which leads to an inhomogeneous broadening in the two-photon resonance. Analogously as in \cite{Rubio'16}, we have rewritten the STIRAP Hamiltonian in the dark/dressed states basis, obtaining the so-called \textit{Optimal Zone} (OZ) of parameters $\Delta^{0}
$, $\Omega_{0}
$, and $v$ for which STIRAP works \cite{Rubio'20}.
In particular, for an atom with velocity $v$ along the propagation direction of the fields, 
this zone is bounded by the curves:
\begin{widetext}
\begin{subequations}
\label{V}
\begin{eqnarray}
\label{Vs}
{\rm Vs^{\pm}}\rightarrow\;\;\;\;\;v 
&=& \cfrac{c}{2\omega_{12}\omega_{13}}\left\{\pm\sqrt{\omega_{13}\left[(\Delta^{0})^2\omega_{13}+\Omega_{0}^2\omega_{12}\right]}-\omega_{13}\Delta^{0}\right\}, \\
\label{Vp}
{\rm Vp^{\pm}}\rightarrow\;\;\;\;\;v 
&=& \cfrac{c}{2\omega_{32}\omega_{13}}\left\{\pm\sqrt{\omega_{13}\left[(\Delta^{0})^2\omega_{13}-\Omega_{0}^2\omega_{32}\right]}-\omega_{13}\Delta^{0}\right\}.
\end{eqnarray}
\end{subequations}
\end{widetext}
Fig.~\ref{f:fig8} shows the contour plot of the numerically calculated population $\rho_{33}$ after the STIRAP process as a function of the atomic velocity $v$, and the ratio $\Delta^{0}/\Omega_{0}$, where the pump and dump pulses coincide with the envelopes of the corresponding trains of pulses used in PAP. We also consider the same atomic frequencies as in the example shown in Fig.~\ref{f:fig8}. The region where the population is transferred to state $\3$ practically coincides with the region enclosed by the curve ${\rm Vs^{\pm}}$ and ${\rm Vp^{\pm}}$, being
\begin{equation}
\label{eq:umbral}
\cfrac{\left|\Delta^{0}\right|}{\Omega_{0}}=\sqrt{\cfrac{\omega_{32}}{\omega_{13}}}, 
\end{equation}
the points at which the $\rm Vp^{\pm}$ curves change from being real to complex valued. Note that these points delimit two zones with different growing behaviors for the base of the curve $\rho_{33}(v)$ as a function of $\Delta^{0}/\Omega_{0}$.
Thus, the FWHM of $\rho_{33}(v)$ will be given by  
\begin{subequations}
\label{eq:Ranges}
\begin{eqnarray}
\label{eq:Ranges1}
W_{ss}&= \frac{1}{2}\left({\rm Vs^{+}-Vs^{-}}\right)\;\;{\rm for}\;\cfrac{\left|\Delta^{0}\right|}{\Omega_{0}}<\sqrt{\cfrac{\omega_{32}}{\omega_{13}}}, \\
\label{eq:Ranges2}
W_{sp}&= \frac{1}{2}\left({\rm Vs^{+}-Vp^{+}}\right)\;\;{\rm for}\;\cfrac{\left|\Delta^{0}\right|}{\Omega_{0}}>\sqrt{\cfrac{\omega_{32}}{\omega_{13}}},
\end{eqnarray}
\end{subequations} 
using that ${\rm Vs^{+}-Vp^{+}}\approx{\rm Vp^{-}-Vs^{-}}$. Arrows in Fig.~\ref{f:fig8} indicate the width of the base of the curve for the parameters used in Fig.~\ref{f:fig9}.
Let us consider that the nominal detuning $\Delta^{0}$ is large enough to neglect the effect of spontaneous emission, which is a requirement to perform our AFC as discussed in the main text. Then, assuming that condition \refeq{eq:Ranges2} is fulfilled, combining it with Eqs.~(\ref{Vs}) and (\ref{Vp}), and using the first order of the Taylor expansion in $\sqrt{1 \pm x}$ with $x=\left|\omega_{12,32}\,\Omega_{0}^2/\omega_{13}(\Delta^{0})^2\right|\ll1$, we obtain that the FWHM of the curve $\rho_{33}(v)$ for which STIRAP is successfully performed reads:
\begin{equation}
\label{eq:FWHMVelocityAP}
W_{sp}=\cfrac{\Omega_{0}^2\,c}{4\omega_{13}\Delta^{0}}.
\end{equation}
In terms of the frequencies, the last expression  corresponds to a FWHM for $\rho_{33}(\delta)$ of 
\begin{equation}
\label{eq:FWHMFrequencyAP}
\varpi_{\rm STIRAP}=\cfrac{\Omega_{0}^2\,\xi}{4\Delta^{0}}, 
\end{equation}
where $\xi=\omega_{34}/\omega_{13}$.

\textit{II. Relation between PAP and STIRAP}.
It has been shown that PAP is equivalent to STIRAP \cite{Shapiro'07} due to the fact that the global evolution operators are the same, as long  as 
each individual pulse of the PAP trains does not significantly change the population. 

As the global evolution of the wavefunction is equal for both processes, we consider that the temporal integrals for all the elements of the density matrix are also the same (we neglect incoherent processes). In particular,
\begin{equation}
\label{eq:c3}
\int_{0}^{t_{f}}\left|c^{\rm PAP}_{3}(t)\right|^{2}dt=\int_{0}^{t_{f}}\left|c^{\rm STIRAP}_{3}(t)\right|^{2}dt,
\end{equation}
where $c_{3}$ is the probability amplitude of state \ket{3}. 
Considering Parseval's theorem about the unitarity of the Fourier transform, we conclude
\begin{equation}
\label{eq:ApA2}
\int_{-\infty}^{+\infty}\!\rho_{33}^{\rm PAP}(\delta)\,d\delta=\int_{-\infty}^{+\infty}\!\rho_{33}^{\rm STIRAP}(\delta)\,d\delta.
\end{equation}
where $\rho_{33}(\delta)=\left|c_{3}(\delta)\right|^2$.

The last expression means that the total area under the curve $\rho_{33}^{\rm STIRAP}(\delta)$ of STIRAP and the sum of the areas of the PAP comb peaks has to be equal.
From \eqref{eq:ApA2} we can obtain a relation between the width $\varpi_{\rm STIRAP}$ and the width of the comb peaks $\varpi_{\rm PAP}$ by proceeding as follows.

First, we assume that the curve $\rho_{33}(\delta)$ for STIRAP and for the $\rm{N_{c}}$ peaks of the PAP comb have Gaussian profiles. Then, we approximate the integral of each Gaussian by the area of a rectangle with the same height as the peak and a base equal to the corresponding FWHM, $\varpi_{\rm STIRAP}$ for STIRAP [see Fig. \ref{f:fig9}(a)] and $\varpi_{\rm PAP}$ for the PAP peaks [see Fig. \ref{f:fig9}(b)]. In the figures, the maximum heights are taken to be 1 for simplicity without lack of generality.
Secondly, we consider the Gaussian envelope of the PAP comb [blue dashed curve in Fig. \ref{f:fig9}(b)].
Using the properties of a Gaussian, it is easy to see that the rectangle $ABCD$, being $\overline{DC}=\sqrt{2\pi}\Gamma$, is divided into two regions with approximately equal areas: the area of the region under the curve, containing the right half of the peaks, and the area of the region over the curve.
This allows us to place the left half of the rectangles, in orange in Fig.~\ref{f:fig9}(b), on top of the right half ones, inside the region $ABCD$ over the curve.
Next, getting rid of the spaces between rectangles, a new one is constructed [see Fig.~\ref{f:fig9}(c)] with a base equal to $({\rm N_{c}}/2)\varpi_{\rm PAP}$, assuming that ${\rm N_{c}\gg1}$.
According to \refeq{eq:ApA2}, the bases of the rectangles of Fig.~\ref{f:fig9}(a) and of Fig.~\ref{f:fig9}(c) must be the same. Therefore, using \eqref{eq:Nc}, we obtain
\begin{equation}
\label{eq:FWHMFrequencyPAP}
\varpi_{\rm PAP}=\cfrac{\sqrt{\pi}\sigma}{T_{int}}\,\varpi_{\rm STIRAP}.
\end{equation}
Finally, casting \eqref{eq:FWHMFrequencyAP} into the last expression, we obtain \eqref{eq:FWHMFrequency}.


\bibliography{BibVC}

\end{document}